\documentclass{aastex631}

\begin{document}

\title{The evolution of CME sheath turbulence from L1 to Earth: Wind and MMS observations of the 2023-04-23 CME}

\correspondingauthor{Matthew R. Argall}

\author[0000-0001-6315-1613]{Matthew R. Argall}
\affiliation{Space Science Center, Institute for the Study of Earth, Oceans, and Space, University of New Hampshire, Durham, NH, USA}

\author[0000-0002-4768-189X]{Li-Jen Chen}
\affiliation{NASA Goddard Space Flight Center, Greenbelt, MD, USA}

\author[0000-0002-1890-6156]{No{\'e} Lugaz}
\affiliation{Space Science Center, Institute for the Study of Earth, Oceans, and Space, University of New Hampshire, Durham, NH, USA}

\author[0000-0001-9210-0284]{Norberto Romanelli}
\affiliation{Department of Astronomy, University of Maryland, College Park, MD, USA}
\affiliation{Planetary Magnetospheres Laboratory, NASA Goddard Space Flight Center, Greenbelt, MD, USA}

\author{Jaye L. Verniero}
\affiliation{Heliospheric Physics Laboratory, NASA Goddard Space Flight Center, Greenbelt, MD, USA}

\author{Charles W. Smith}
\affiliation{Space Science Center, Institute for the Study of Earth, Oceans, and Space, University of New Hampshire, Durham, NH, USA}

\author[0000-0001-8702-5806]{Brandon Burkholder}
\affiliation{NASA Goddard Space Flight Center, Greenbelt, MD, USA}

\author[0000-0002-2463-4716]{Victoria Wilder}
\affiliation{University of Texas at Arlington, Arlington, TX, USA}

\begin{abstract}
An interplanetary shock driven by a coronal mass ejection (CME) containing an interval of sub-Alfv\'enic flow impacted Earth on April 23, 2024. In this article, we analyze the turbulence in the sheath region between the shock and CME to determine how it evolves from L1 (as observed by Wind) to Earth (as observed by MMS, upstream of the bow shock). Wind and MMS were separated by $55\,\mathrm{R_{E}}$ in the dawn-dusk direction, but the shock normals differ by only $2.8^{\circ}$ and the Pearson correlation coefficient between time-shifted magnetic field components is $\rho=0.93$. We observe a shift in the break point of the magnetic power spectral density between inertial and ion kinetic scales toward the ion inertial length and a steepening of the spectral slope, indicating more active energy cascade closer to Earth. The distribution of increments becomes more non-Gaussian near Earth, particularly at ion kinetic scales, indicating the turbulence becomes more intermittent. Finally, the correlation length at Earth is 25\% longer than at L1, indicating that the turbulence is smoothing out the magnetic field. The results present an example of substantial evolution of CME sheath turbulence from L1 to Earth.
\end{abstract}




\section{Introduction}
\label{sec:intro}
Coronal mass ejections (CMEs) from the Sun are key drivers of the space weather at Earth \citep{Gosling:1993, Temmer:2023}. When a CME moves faster than the ambient solar wind, an interplanetary forward fast magnetosonic shock develops. The region between the CME and the shock, called the CME sheath, is turbulent and can contain energetic particles \citep{Lario:2002, Lario:2023, Kilpua:2023} and large-scale planar structures \citep{Kilpua:2017,Nakagawa:1989}. Some studies on CME sheath turbulence compare fluctuations in the upstream solar wind to fluctuations in the sheath \citep{Good:2020}. Other studies differentiate between the post-shock and pre-CME regions of the sheath \citep{Kilpua:2021, Kilpua:2020}. Fewer address the radial evolution of the CME sheath turbulence \citep{Manoharan:2000, Good:2020}. In this study, we compare CME sheath turbulence from the Lagrange point 1 (L1) to Earth. Generally, L1 and Earth are considered to be the same location on interplanetary scales \citep[with some exceptions for IMF clock angle][]{Walsh:2019,Borovsky:2021}, and as such, the evolution of sheath turbulence from L1 to Earth has been overlooked. In this paper, we show that within this short interplanetary distance, CME sheath turbulence evolves significantly, providing new insight into turbulence dynamics and CME sheaths.

At 1 AU, CME sheath dynamics are often compared to the upstream solar wind, and subregions of the sheath with different properties have also been identified, namely the region near the shock and near the magnetic ejecta \citep{Kilpua:2020}. The CME sheath typically contains 10 times more energy in the magnetic field than the upstream solar wind. Energy in the inertial range remains steady until 10 nonlinear times (a measure of the eddy turnover time scale) but decreases immediately in the kinetic range, suggesting the shock provides energy at large scales but equilibrium is initiated via energy transfer at small scales \citep{Pitna:2017}. The magnetic field power spectral index is near the Kolmogorov scaling of $-5/3$ in the inertial range \citep{Kilpua:2021, Pitna:2017, Good:2020, Bourouaine:2012} and steepens to $-2.8$ at ion kinetic scales \citep{Good:2020}. However, the kinetic range magnetic spectra can be steeper near the ejecta leading edge than directly behind the shock, possibly due to the formation of compressible structures like current sheets \citep{Kilpua:2021}. That said, increments of the magnetic field \citep[the difference between lagged measurements; see Section~\ref{sec:methods} or][]{Greco:2008} are largest near the shock, smaller near the ejecta leading edge, and smallest in the solar wind \citep{Kilpua:2021}. Compressibility is larger in the sheath (comparable to the slow solar wind) than in the upstream solar wind, as is the angular deflection of the magnetic field associated with each increment. Past studies have almost always relied on measurements from L1 because of the continuous coverage there. To see how the sheath evolves as it propagates Earthward, observations from radially aligned spacecraft dispersed throughout interplanetary space are required.

One of the few such studies relying on radial conjunctions studied the evolution of a CME sheath region from the MErcury Surface, Space ENvironment, GEochemistry, and Ranging (MESSENGER) mission at 0.47\,AU and STEREO-B at 1.08\,AU \citep{Good:2020}. Probability distributions of $\Delta B/B$ showed large tails in the sheath at both radial locations, indicating the presence of intermittent turbulent structures, but the tails were absent in the solar wind upstream of the sheath at MESSENGER. Magnetic fluctuations in the sheath generally became more compressive with radial distance (clustered near the sheath trailing edge). At 1.08\,AU, the sheath was relatively less compressive (containing enhanced Alfv\'enic fluctuations) compared to the upstream solar wind than at 0.47\,AU. In addition, the inertial range spectral slope steepened, sheath complexity decreased (but increased relative to upstream solar wind), and planar structures grew in duration \citep{Good:2020}.

In this paper, we analyze the radial evolution from L1 to Earth of the magnetic turbulence within the sheath of a CME that arrived at Earth on 2023 April 23. The CME drove an interplanetary shockand resulted in a geomagnetic storm. The associated magnetic cloud contained an interval of sub-Alvf\'enic flow that caused the bow shock and magnetotail to dissipate \citep{ChenLJ:2024, Burkholder:2024}. The magnetosphere developed Alfv\'en wings that connect to the solar corona \citep{Gurram:2025}. We focus on the turbulent region in the CME sheath region just behind the shock and show that despite L1 being just 0.01 AU upstream from Earth, the turbulence characteristics can change dramatically over such small distances.

\section{Data}
\label{sec:data}
We use data from the Wind \citep{Ogilvie:1997} and Magnetospheric Multiscale (MMS) missions \citep{Burch:2015}.
Specifically, we use data from their magnetic field and ion plasma instruments to study the evolution of magnetic turbulence in a CME sheath from L1 to Earth.

The Wind mission \citep{Ogilvie:1997, WilsonIII:2021} was launched on 1 Nov. 1994 into an Earth-centric orbit with apogee at 250 Earth radii ($R_{E}$) then was eventually inserted at the Sun-Earth Lagrange 1 (L1) point in 2004 to serve as an upstream solar wind monitor. At the time of the CME, it was located at ($200$, $36$,  $-10$)\,$R_{E}$ in Geocentric Solar Ecliptic (GSE) coordinates. Magnetic field data from Wind is recorded by the Magnetic Field Investigation \citep[MFI;][]{Lepping:1995}, which samples at $10.9\,s^{-1}$, and plasma moments are provided by the Solar Wind Experiment \citep[SWE;][]{Ogilvie:1995} once every 100\,s.

\begin{figure}
\plotone{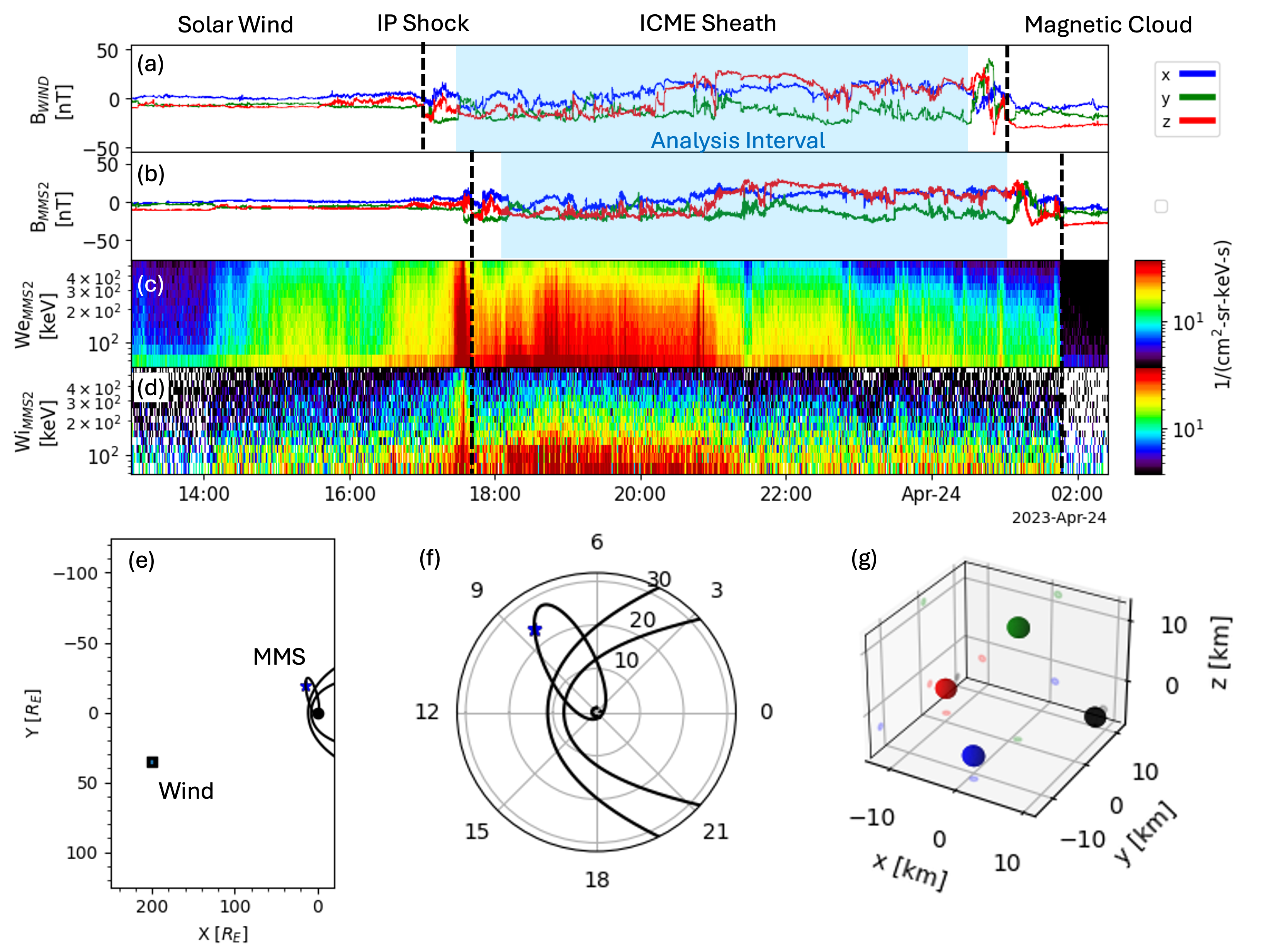}
\caption{An interplanetary shock driven by a CME arrived at L1 at 16:54\,UT on 2023-04-23 then at MMS 40 minutes later. The magnetic field from (a) Wind (b) and MMS, and the energetic (c) electrons and (d) ions from MMS show the CME sheath as a turbulent region filled with energetic particles. (e) Wind was duskward of the Earth-Sun line by $36\,R_{E}$ while MMS dawnward of the Earth-Sun line by $19\,R_{E}$. (f) MMS was located in the solar wind ${\approx}10\,\mathrm{R_{E}}$ normal to the bowshock (f) with an inter-spacecraft separation of ${\sim}15\,\mathrm{km}$
\label{fig:overview}}
\end{figure}

The MMS mission was launched on 15 Mar. 2015 into a $1.08 \times 12\,R_{E}$ orbit. Apogee was eventually raised to ${\sim}29\,R_{E}$, allowing MMS to consistently sample the magnetosheath, bow shock, and solar wind while on the day-side of Earth. During the analysis interval, MMS was located at ($14$, $-19$,  $-10$)\,$R_{E}$, approximately $10\,R_{E}$ ($14.7\,R_{E}$) from the model bow shock (magnetopause), had a mean inter-spacecraft separation of $15$\,km (Fig.~\ref{fig:overview}f), and was initially in slow survey mode \citep{Fuselier:2016}. The DC component of the magnetic field from $0-8$\,Hz is measured by the Fluxgate Magnetometer \citep[FGM;][]{Russell:2014} while the AC component, from $0.3-16$\,Hz is measured by the the Search Coil Magnetometer \citep[SCM;][]{LeContel:2014}. Plasma moments are measured by the Fast Plasma Investigation \citep[FPI;][]{Pollock:2016} after MMS transitions to fast survey at 23:23\,UT on April 23 and are sampled once every $4.5$\,s. The non-thermal plasma is sampled by the Fly's Eye Energetic Particles Spectrometers \citep[FEEPS;][]{Blake:2016} once every 20\,s in fast survey and once every 200\,s in slow survey mode.

\section{Methodology}
\label{sec:methods}
We examine the power spectral density (PSD), distribution of increments, and correlation length of the magnetic field in the solar wind at L1 and near Earth, just upstream of the bow shock, to determine how turbulence in the CME sheath evolves. The reduced PSD of the magnetic field shows how much energy is contained in the magnetic field as a function of frequency (or scale length). The slope of the PSD indicates the rate at which energy is being transferred to the plasma, and spectral breakpoints at which the slope changes denote scales at which the energy transfer mechanism changes. The reduced PSD $\mathbf{B}$ is defined as
\begin{equation}
    P(k) = \frac{1}{3} P_{i,i}(k),
\end{equation}
where $P_{i,j}$ is the cross-spectral matrix of the magnetic field, $i,j \in (x, y, z)$ are the vector components in GSE coordinates, repeated indices indicate Einstein summation notation, $k = f/\left<V_{i}\right>$ represents the wave number, $f$ is frequency, and $\left<V_{i}\right>$ is the time-averaged ion bulk velocity. In converting from frequency to wave number space, we made use of the Taylor approximation, which assumes that the temporal variations are much slower than spatial variations convecting past the spacecraft. From the PSD, spectral indices $\alpha$ are determined by fitting a power law $P = k^{\alpha}$ to sliding one-decade windows. 

Increments of the magnetic field can be calculated from single-spacecraft and from spacecraft pairs. Single-spacecraft increments are
\begin{equation}
    \Delta B_{i}^{l}(t) = B_{i}(t+\tau) - B_{i}(t),
\end{equation}
where $l = \left<V_{i}\right> \tau$ is the spatial lag determined from temporal lag $\tau$ assuming the Taylor approximation. Pair-wise increments between spacecraft $a$ and $b$ are
\begin{equation}
    \Delta B_{i}^{ab}(r_{ab}) = B_{i}^{a}(t) - B_{i}^{b}(t).
\end{equation}
From these increments, we compute the equivalent spectrum for multi-spacecraft validation of single-spacecraft PSD estimates (Eqs.~\ref{eq:multisc-struct-fn} and \ref{eq:singlesc-struct-fn}, respectively), and determine the intermittency of the fluctuations by computing the probability density function (PDF) of the increments.

Estimates of the PSD, the equivalent spectra, are calculated from the second-order structure function \citep[see][and references therein]{Chhiber:2018} using spacecraft pairs
\begin{equation}
    \lambda D_{i}^{(2)}(r_{ab}) = \left< \left[ \Delta B^{ab}_{i} (t) \right]^{2} \right>
    \label{eq:multisc-struct-fn}
\end{equation}
and single-spacecraft time series
\begin{equation}
    \lambda D_{i}^{(2)}(l) = \left< \left[ \Delta B^{l}_{i} (t) \right]^{2} \right>.
    \label{eq:singlesc-struct-fn}
\end{equation}
Here, $\lambda$ is the effective wavelength associated with lag $r_{ab}$, which, for two-spacecraft estimates, is the instantaneous spatial lag between spacecraft $r_{ab} = |\mathbf{r}_{a} - \mathbf{r}_{b}|$, and for single-spacecraft estimates is the time-delayed lag $l = \left< V_{i} \right> \tau$.

Intermittency refers to a sporadic, bursty dissipation of energy in localized enhancements \citep{Bruno:2013}. It can indicate the presence of small-scale reconnecting current sheets and local plasma acceleration \citep{Matthaeus:2011}, produce non-uniform plasma heating \citep{Alexandrova:2008}, and modulate solar wind electric fields for more efficient solar wind-magnetosphere coupling \citep{Zhdankin:2015}. Intermittency is determined from the PDF of 
$(\Delta B_{i}^{l}(t) - \mu)/\sigma$, where $\mu$ and $\sigma$ are the mean and standard deviation of $\Delta B_{i}^{l}(t)$, respectively. This allows us to compare the PDF to a standard normal Gaussian distribution, $\mathcal{N}$. The kurtosis of the PDF
\begin{equation}
    \kappa = \int^{\infty}_{-\infty} \left( \frac{\Delta B_{i}^{l}(t) - \mu} {\sigma}  \right)^{4} f\left[ \Delta B_{i}^{l}(t) \right] dt
\end{equation}
tells us if it is sub-Gaussian ($\kappa < 3$) or super-Gaussian ($\kappa > 3$). A super-Gaussian distribution has large tails, indicating intermittent turbulence with sporadic, large amplitude fluctuations.

The correlation length indicates the size of the energy containing scale at which large turbulent eddies exist before being broken down into smaller eddies in the inertial range and below \citep{Matthaeus:1999, Smith:2001}. This scale is important for describing field-line random walk and charged particle diffusion and describes the separation over which parcels of magnetic field remain similar \citep{Matthaeus:1999}. More formally, the correlation length, or e-folding scale, is the length scale at which the auto-correlation coefficient at a given lag of the magnetic field decreases by a factor of $e$ (i.e. becomes small). The autocorrelation of the magnetic field is
\begin{equation}
    R(\tau) = E[(X_{t} - \mu_{x}^{\prime}) (Y_{t-\tau} - \mu_{y}^{\prime})]
\end{equation}
where the expectation value $E[X]$ and the local means $\mu^{\prime}_{x}$, $\mu^{\prime}_{y}$  are defined by
\begin{eqnarray}
	E[X] &= \frac{1}{N-\tau} \Sigma_{t=\tau}^{N} X_{t} \\
	\mu_{x}^{\prime} &= \frac{1}{N-\tau} \Sigma_{t=\tau}^{N} X_{t} \\
	\mu_{y}^{\prime} &= \frac{1}{N-\tau} \Sigma_{t=\tau}^{N} Y_{t-\tau}.
\end{eqnarray}
In these equations, $t=1, 2, \dots, N$, $N$ is the sample size, and $\tau$ is the lag. By using the local mean $\mu^{\prime}$ instead of the global mean $\mu_{x} = N^{-1} \Sigma^{N}_{t=1} X_{t}$, as is typically done for covariance operations, the correlation function $R(\tau)$ is locally detrended, a common operation used to prevent leakage of power at low wavenumber from spreading throughout the spectrum \citep{Matthaeus:1982, Bandyopadhyay:2020}.

\begin{table}
    \centering
    \begin{tabular}{cccccccccc}
        S/C  & $t_{0}$  & $t_{1}$  & $f_{s}$ [S/s] & $V_{i}$ [km/s] & $\Delta x$ [km] & $\rho_{i}$ [km] & $d_{i}$ [km] & $\rho_{e}$ [km] & $d_{e}$ [km] \\
        Wind & 17:15:00 & 00:45:00 & 11 & 619.2 & 56.3 & 6074.2 & 60.4 & 3.3 & 1.4 \\
        MMS & 17:40:00 & 01:10:00 & 16/32 & 637.0 & 39.8/19.9 & 6082.0 & 55.7 & 3.3 & 1.3 \\
    \end{tabular}
    \caption{Plasma parameters observed by Wind at L1 and MMS upstream from the bowshock. ($t_{0}$, $t_{1}$) is the analysis interval and interval over which other quantities were averaged; $f_{0}$ is the sample rate of the magnetometer (FGM/SCM for MMS); $V_{i}$ is the magnitude of the ion velocity; $\Delta x = V_{i}/f_{0}$ is the propagation distance of the solar wind in one sample time; $\rho_{i}$ ($\rho_{e}$ is the ion (electron) larmour radius; and $d_{i}$ ($d_{e}$) is the ion (electron) inertial length.}
    \label{tab:plasma-params}
\end{table}

\section{Results}
\label{sec:results}
An interplanetary shock driven by a CME arrived at L1 and was observed by the Wind spacecraft at approximately 16:54\,UT on 2023-04-23 (Fig.~\ref{fig:overview}a). 40 minutes and 35 seconds later, as determined by a correlation analysis \citep[e.g.,][]{Case:2012}, the shock arrived upstream of Earth's bow shock and was observed by MMS (Fig.~\ref{fig:overview}b-d) which was in the solar wind ${\sim}10\,\mathrm{R_{E}}$ upstream of the bow shock along the bow shock normal (Fig.~\ref{fig:overview}e,f). The shock is identified as an abrupt change in the magnetic field components (Fig.~\ref{fig:overview}a,b; first vertical dashed line) preceded by a significant increase in energetic ions (Fig.~\ref{fig:overview}c) and electrons (Fig.~\ref{fig:overview}d). The shock at both spacecraft is trailed by a turbulent sheath region that is filled with high energy particles, particularly in the first half of the interval. After about 8 hours, the spacecraft exit the sheath and enter the CME magnetic cloud, marked by a more steady magnetic field and abrupt cut-off of energetic particles (Fig.~\ref{fig:overview}a-d; second vertical dashed line). Despite Wind and MMS being separated by $55\,\mathrm{R_{E}}$ in the $y$-GSE direction, a minimum variance analysis of the magnetic field results in shock normals of $\hat{n}_{WIND} = (0.8534, 0.2354, -0.4650)$ and $\hat{n}_{MMS} = (0.8357, 0.2800, -0.4725)$, a difference of only $2.8^{\circ}$. In addition, the Pearson correlation coefficient of the time-shifted magnetic field components is $\rho=0.93$. Together, this indicates that the spacecraft crossed the shock at similar locations and are observing regions of similar magnetic topology.

For more details on this event, particularly the magnetic cloud and its geomagnetic impact, the reader is referred to \citet[][and references there in]{ChenLJ:2024}. We focus on the sheath region and analyze turbulent properties that may be responsible for producing the energetic particles.

\begin{figure}
\plotone{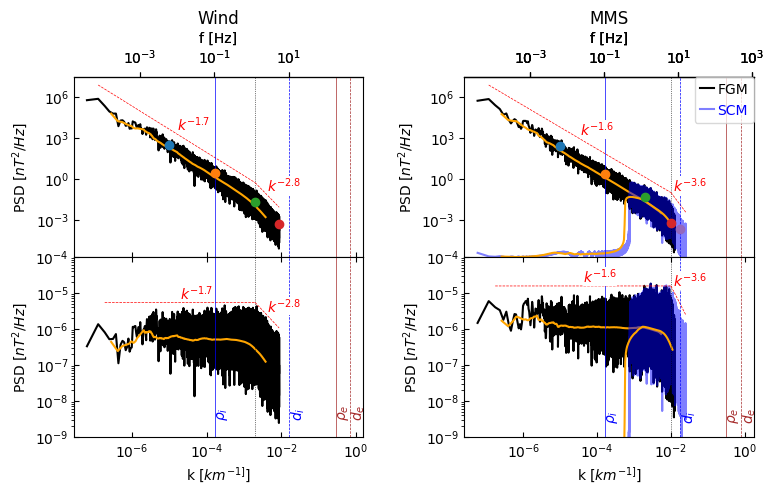}
\caption{The ion-kinetic scale breakpoint shifts toward $d_{i}$ from L1 (left) to Earth (right) and the spectral slope steepens. The reduced PSD (black) is computed during the CME sheath interval, with the median-average shown in yellow. The ion (electron) larmor radius and inertial lengths are shown as blue (brown) solid and dashed lines, respectively. The spectral break is indicated by a vertical dotted black line. Spectral slopes in the inertial and ion-kinetic ranges are shown as red dashed lines and were determined from fitting the spectra with sliding windows of one decade. Colored dots correspond to the scales at which the increments in Figure~\ref{fig:intermittency-xyz} were calculated. In the bottom panels, the PSD is scaled by the inertial range spectral index to highlight goodness of fit and enhance the spectr break.
\label{fig:psd}}
\end{figure}

\subsection{Spectral Slopes \& Breakpoints}
\label{sec:power}
Figure~\ref{fig:psd} (top) shows the reduced power spectral density (PSD) of the magnetic field at Wind (left) for the time interval 17:15:00 to 00:45:00\,UT and at MMS (right) for the time interval 17:40:00 to 01:10:00\,UT (see also the blue shaded regions in Fig.~\ref{fig:overview}). Frequency (upper $x$-axis) was converted to wave number space (lower $x$-axis) using the Taylor approximation, $k = f/\left< V \right>_{i}$, where $\left< V \right>_{i}$ is the average ion bulk velocity during the interval. Average plasma parameters for Wind and MMS are found in Table~\ref{tab:plasma-params}. The gyroradius (solid; $\rho_{s} = \frac{m_{s} v_{\perp,s}} {Z_{s} e |B|}$, where $m_{s}$ is the mass of species $s$, $v_{\perp,s}$ is the magnitude of the bulk velocity perpendicular to the magnetic field $B$, $Z_{s}$ is the charge number, and $e$ is the elementary charge) and the inertial length (dashed; $d_{s} = c / \omega_{p,s}$, where $c$ is the speed of light, $\omega_{p,s} = \sqrt{\frac{e_{s}^{2} n_{s}}{\epsilon_{0} m_{s}}}$ is the plasma frequency, $n_{s}$ is the number density, and $\epsilon_{0}$ is the permittivity of free space) are shown for ions (blue) and electrons (brown). The gold trace depicts a sliding half-decade median average. The spectra exhibit a break point followed by a steeper slope between the ion gyroradius and inertial lengths.

The spectral break points shift to larger wave number from Wind to MMS. The spectra exhibits power laws of ${\approx}k^{-5/3}$ at fluid scales at both L1 and Earth (red dashed line), consistent with MHD turbulence \citep{Kolmogorov:1991}. The bottom panels show the PSD scaled by the fluid-scale power law so that the spectra in that range has a slope of 1.  Wind at L1 observes a break in the spectra at $k=2\times10^{-3}\,\mathrm{km}^{-1} \approx 8.3\,d_{i}$ (black dashed line). Downstream at Earth, MMS observes a spectral break at $k=1\times10^{-2}\,\mathrm{km}^{-1} \approx 1.8\,d_{i}$. Spectral breaks often occur at ion kinetic scales, where magnetic fluctuations become comparable in size to characteristic ion scale lengths. This can be the result of energy conversion, dissipation, and dispersion processes \citep{Bourouaine:2012, Smith:2011, Markovskii:2008}. A shift in the spectral break toward the ion inertial length could signal a change in how energy transfer is occurring.

At scale sizes smaller than the spectral break, the PSD steepens, indicating an increased energy conversion rate from the magnetic field. At L1, the power law becomes $k^{-2.8}$ while at Earth it steepens to $k^{-3.6}$. Furthermore, at Earth, there is more energy at scale sizes between the Wind and MMS breakpoints (Fig.~\ref{fig:psd}), indicating that some of the energy may have cascaded back to larger scales.

\begin{figure}
    \includegraphics[width=0.46\textwidth]{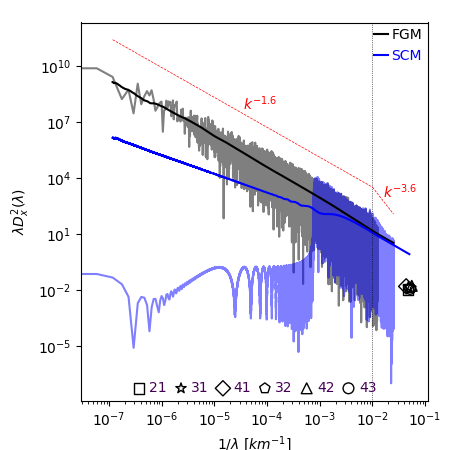}
    \caption{Multi estimates of the equivalent spectra from each of the six spacecraft pairs (shapes with numbers identifying the spacecraft pairs -- 21 is for the MMS2-MMS1 pair, etc.) are consistent with spectral break point occurring near the ion inertial length in the single-spacecraft estimates of the PSD (semi-transparent lines) for FGM (black) and SCM (blue). The single-spacecraft estimates of the equivalent spectrum (solid lines) do not show a spectral break, suggesting a difference along the spacecraft separation baseline and bulk flow directions.
    \label{fig:struct-func}}
\end{figure}

At MMS, the spectral break is near the Nyquist frequency of both the FGM and SCM instruments, so aliasing or downsampling may affect the spectral break point. Fortunately, the spacecraft separation is $k \sim 0.07\,\mathrm{km}^{-1}$ so multi-spacecraft methods can probe scales smaller than those accessible to single spacecraft. Figure~\ref{fig:struct-func} shows the equivalent spectrum (Eqs.~\ref{eq:singlesc-struct-fn}) from MMS2 (solid) superimposed on the power spectral density (semi-transparent), as well as the multi-spacecraft equivalent spectrum calculated from each spacecraft pair (symbols: Eq.~\ref{eq:multisc-struct-fn}). The power spectral density has been shifted up to align with the equivalent spectrum. At frequencies near and below 0.3\,Hz, the SCM equivalent spectrum is affected by the high-pass filter applied to survey data and visible in the PSD. Above 0.3\,Hz, the FGM and SCM equivalent spectrum agree but do not show the spectral break present in power spectral density. However, the multi-spacecraft equivalent spectra do agree with the spectral break and steepened slope. Discrepancies between the single- and multi-spacecraft methods can result from the fact that the Taylor hypothesis assumes variations along the radial direction while the two-spacecraft method identifies variations along the inter-spacecraft baseline \citep{Chhiber:2018, Chasapis:2020}. Overall, the two-spacecraft equivalent spectra is consistent with the change in slope observed in the PSD.

\begin{figure}
    \includegraphics[width=\textwidth]{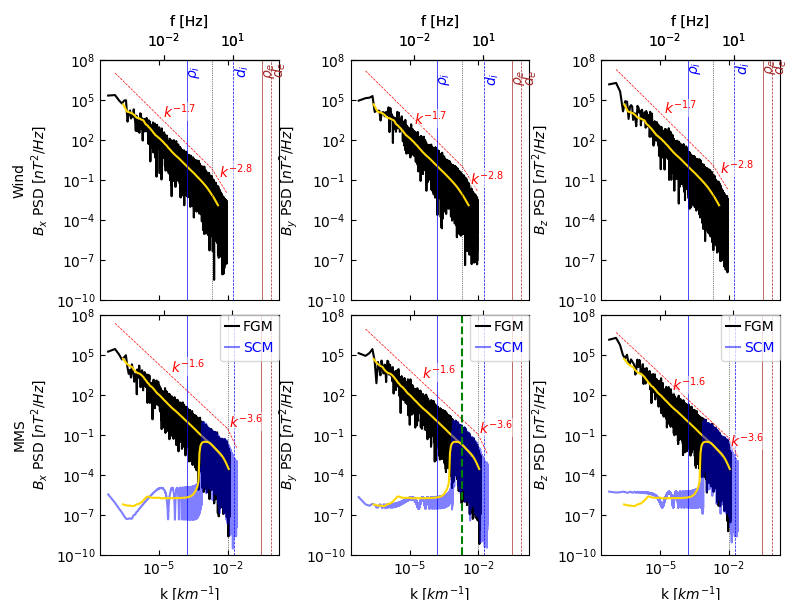}
    \caption{PSD for $B_{x}$ (left) and $B_{z}$ (right) at Wind (top) and MMS (bottom) consistent with the omni-directional PSD shown in Figure~\ref{fig:psd}. However, $B_{y}$ (center) at MMS does not exhibit a shift in spectral break point.
    \label{fig:psd-xyz}}
\end{figure}

Figure~\ref{fig:psd-xyz} shows the PSD of each component of the magnetic field at Wind (top) and MMS (bottom) in the same format as the top panels of Figure~\ref{fig:psd}. At L1, all components are consistent with the reduced PSD. At Earth, however, only $B_{x}$ and $B_{z}$ match the reduced PSD. The spectral break at $B_{y}$ is more consistent with the break point observed at L1 (vertical green dashed line), instead of that observed in the other two components by MMS (black vertical dotted line), indicating the turbulence develops anisotropically.

\subsection{Intermittent Structures}
\label{sec:intermittency}
\begin{figure}
    \plotone{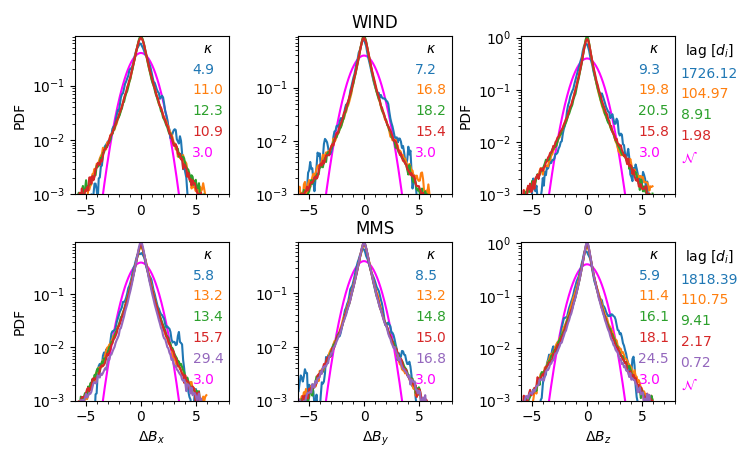}
    \caption{PDFs of the standard normal increments for $B_{x}$ (left), $B_{y}$ (center) and $B_{z}$ (right) at L1 (top) and Earth (bottom) calculated at inertial (blue), ion (orange, green, red), and sub-ion (purple) scales. Increments were calculated at lags corresponding to the colored dots in Figure~\ref{fig:psd}. A kurtosis $\kappa > 3$ indicate wider tails than the Gaussian distribution $\mathcal{N}$ (magenta), indicating turbulence is intermittent. At L1, the intermittency (kurtosis, $\kappa$) peaks between $\rho_{i}$ (orange) and $d_{i}$ (red) scales. At Earth, intermittency increases monotonically and increases significantly at sub-ion scales. Increased intermittency is associated with a lower probability of intermediate-sized fluctuations ($2 < \Delta B_{i} < 4$).
    \label{fig:intermittency-xyz}}
\end{figure}

Figure~\ref{fig:intermittency-xyz} shows probability density functions (PDFs) of the standard normal increments of $B_{x}$ (left), $B_{y}$ (center), and $B_{z}$ (right) at different lags, assuming the Taylor hypothesis. For Wind (top), the lags were chosen to be [$1\times10^5$\,km, $\rho_{i,\mathrm{Wind}}$, $1/k_{\mathrm{break},\mathrm{Wind}}$, $1/k_{\mathrm{Nyquist},\mathrm{Wind}}$] while for MMS (bottom), the lags are [$1\times10^5$\,km, $\rho_{i,\mathrm{MMS}}$, $1/k_{\mathrm{break},\mathrm{Wind}}$, $1/k_{\mathrm{break},\mathrm{MMS}}$, $d_{i,\mathrm{MMS}}$]. Here, note that $d_{i,\mathrm{Wind}}$ is slightly larger than $d_{i,\mathrm{MMS}}$, $k_{\mathrm{break}}$ is the location of the spectral break, and color-coded lags correspond to the dots on the PSD curves in Figure~\ref{fig:psd}. The kurtosis $\kappa$ is computed for each PDF to indicate how (non-)Gaussian the distributions of increments are. In general, $\kappa$ is larger at ion kinetic scales $\rho_{i}^{-1} \le k < d_{i}^{-1}$ than it is at inertial scales $k < \rho_{i}^{-1}$, indicating an excess of large-amplitude structures at ion scales.


MMS is able to sample at scales slightly less than the ion inertial length. Beyond ion kinetic scales, $\kappa$ increases drastically, with the PDF becoming more peaked through the loss of intermediate-amplitude fluctuations in the range $2 < |(\Delta B - \mu)/\sigma| < 4$. Since Wind does not see a similar increase in kurtosis beyond the spectral break, the change in the distribution of increments at MMS, particularly loss of intermediate-amplitude increments, may again be related to a change in dissipation mechanisms and increase energy conversion rates.

From L1 to Earth, the distributions of increments are persistent while showing signs of energy loss. At sub-ion scales (blue curve), a large bump on-tail can be seen at both Wind and MMS. These are due to the gradual, somewhat linear changes in the magnetic field components [e.g., $B_{x}$ from 18:00 (18:30)\,UT and $B_{y}$ and $B_{z}$ from 21:00 (21:45)\,UT at L1 (Earth)]. At both sub-ion and ion scales, the distributions have similar shapes and values of kurtosis; however, $\kappa$ increases slightly from L1 to Earth. At Earth, the largest amplitude fluctuations that make up the tail of the distribution remain equally probable while intermediate fluctuations are less probable, making the distribution more peaked and the tails more pronounced. As a result, $\kappa$ increases from ${\sim}11$ to ${\sim}14$. From L1 to Earth, then, sub-ion-scale features persist while ion-scale fluctuations of intermediate amplitude are less probable.

At L1, $B_{y}$ and $B_{z}$ are slightly more intermittent than $B_{x}$, but the general trends are similar. At ion scales, there is an increase in kurtosis mainly due to a decrease in probability of fluctuations $2 < |(\Delta B_{i} - \mu)/\sigma| < 4$ rather than an increase in probability of large amplitude fluctuations. At Earth, the fluid- and ion-scale PDFs are similar between the three components; however, at sub-ion scales $B_{y}$ does not exhibit the significant increase in kurtosis observed in the other two components. This is consistent with the spectral break in the PSD of $B_{y}$ not changing between L1 and Earth (Fig.~\ref{fig:psd-xyz}) and again indicates that the turbulence develops anisotropically.

\begin{figure}
    \plotone{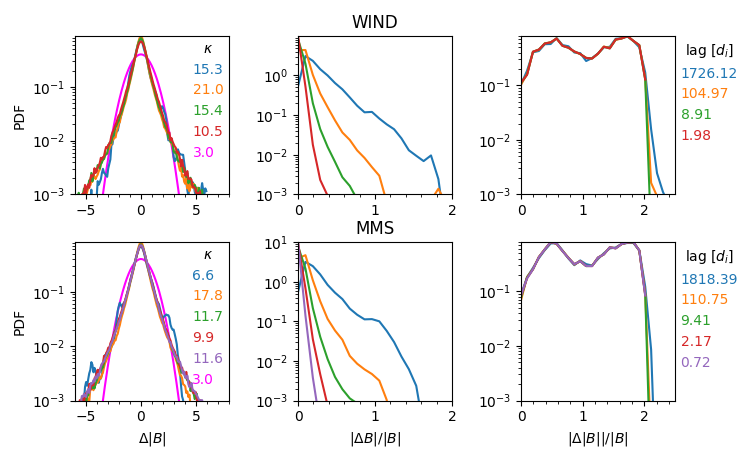}
    \caption{Standard normal increments of $|B|$ (left) at L1 (top) and Earth (bottom) show that intermittency peaks in ion kinetic scales (orange). Normalized increments (center) become progressively more Gaussian with decreasing scale size. Compressibility (right) has a sharp cut-off at 2, indicating the fluctuations mostly Alfv\'enic. 
    \label{fig:compressibility}}
\end{figure}

For insight into the type of fluctuations occurring in the CME sheath, we present the PDF of the normalized fluctuation amplitudes in Figure~\ref{fig:compressibility} for Wind (top) and MMS (bottom), including the standard normal increments, $\Delta |B|$ (left), the normalized increments, $|\Delta B^{l}(t)|/|B|$ (center), and the magnetic compressibility, $|\Delta |B^{l}(t)||/|B|$ (right). Here, $|B|$ is the mean-field magnitude between times $t$ and $t+\tau$ that define the scale size $l$. $\Delta |B|$ is more intermittent at fluid scales at both L1 and Earth than the individual components. The kurtosis of $\Delta |B|$ is notably higher at L1 and is observed to increase at ion gyro-radius scales and decrease toward the ion inertial length. Similar trends are observed at Earth, but while the components exhibit a large increase in intermittency at sub-ion scales, no such increase is observed in $\Delta |B|$. The normalized increments are sharply and narrowly peaked at the smallest scales, indicating weak fluctuations with respect to the mean field, and gradually get more broad with increasing scale size, a feature typical of the solar wind \citep{Kilpua:2021}. At fluid scales, $|\Delta B^{l}(t)|/|B|$ is larger at L1 than at Earth. The compressibility shows that $|\Delta |B^{l}(t)||/|B| {<\sim} 2$ at all scales. For a pure rotation, the largest change in magnitude would be twice the mean field, meaning the fluctuations are predominantly Alfv\'enic in nature.

\subsection{Correlation Length}
\label{sec:correlation}

Figure~\ref{fig:corr-xyz} shows the evolution of the auto-correlation of $B_{x}$ (left), $B_{y}$ (center), and $B_{z}$ (right) as a function of lag from L1 (top) to Earth (bottom). The lag at which the correlation coefficient becomes small ($1/e$) indicates the correlation length scale, $\lambda_{C}$. The correlation length has been determined in two ways: first, by finding the lag at which the auto-correlation $R(r)/R(0) < 1/e$ such that the exponent in $y = e^{x/\lambda_{C}}$ is less than one; and second, by fitting the curves with the function $y = ae^{-x/\lambda_{C}}$ (green; $a=1$ orange). The result is that the correlation length of $B_{x}$ increases from $\lambda_{C} \approx 4\times10^{4}$\,km at L1 to $\lambda_{C} \approx 5\times10^{4}$\,km at Earth. This is consistent with continued MHD dissipation rates observed in Figure~\ref{fig:psd} and the increase in kurtosis in Figure~\ref{fig:intermittency-xyz}. Turbulence is removing energy from the magnetic field, decreasing the probability of intermediate-amplitude fluctuations, effectively smoothing out the magnetic field. As a result, the correlation length increases because each parcel of magnetic field looks more similar to its neighbor.


The correlation length for both $B_{y}$ and $B_{z}$ also increase from L1 to Earth. That of $B_{y}$ is notably shorter than the other two components and the correlation function matches the frozen-in condition nicely. This is consistent with both the increments (Fig.~\ref{fig:intermittency-xyz}) and the PSD (Fig.~\ref{fig:psd-xyz}) for $B_{y}$ not evolving much between L1 and Earth.

\section{Discussion}
\label{subsec:discussion}
In this paper, we analyzed the evolution of CME sheath turbulence from L1 to Earth. In the literature, it is more common to see a comparison of the CME sheath turbulence with the upstream solar wind \citep{Kilpua:2021}, of different regions within the sheath \citep{Pitna:2017, Kilpua:2021}, or of sheath properties over much larger heliocentric distances \citep{Bourouaine:2012, Good:2020}. Figure~\ref{fig:overview} shows a distinct decrease in energetic ions and electrons about mid-way through the sheath region, near 21:15\,UT, suggesting that the post-shock and pre-ejecta regions in the sheath are subject to different influences; the post-shock region likely contains shocked material while the pre-ejecta material is likely swept up material of coronal origin \citep[see, e.g.,][]{Kilpua:2017}. A comparison of these two regions for this event is worthy of a follow-up study. We compared the sheath regions between L1 and Earth because the two locations are often thought of as coincident on interplanetary scales, because of the geoeffectiveness that CMEs have and the role they play in our local space weather, and because there is open debate on how well upstream data can predict conditions at Earth. We showed that the turbulence within the CME sheath region indeed evolves from L1 to Earth.

\begin{figure}[ht!]
    \plotone{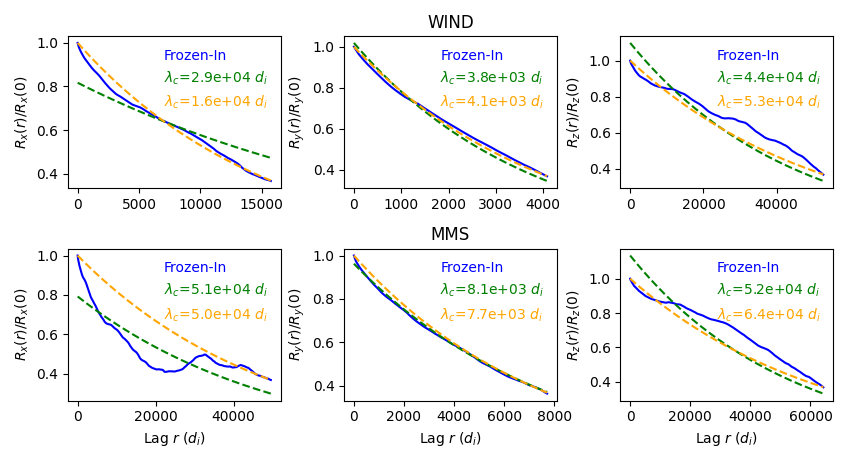}
    \caption{Correlation lengths at L1 (top) and Earth (bottom) for $B_{x}$ (left), $B_{y}$ (center), and $B_{z}$ (right). Generally, $\lambda_{C}$ increases from L1 to Earth. This is consistent with energy cascade from large to small scales, making adjacent parcels of magnetic field appear more similar at coarse scales. For $B_{y}$, $\lambda_{C}$ is notably shorter and follows the frozen-in assumption. Frozen-in assumption fits the data (blue) to $a e^{-r/\Lambda_{C}}$ (green) and $e^{-r/\Lambda_{C}}$ (orange).}
    \label{fig:corr-xyz}
\end{figure}

A spectral slope of $k^{-5/3}$ that follows the Kolmogorov scaling in the inertial range, and one of $k^{-2.8}$ in the kinetic range appear to be universal features of both the fast and slow solar wind \citep{Alexandrova:2009, Bruno:2017}. Several unrelated fast and slow solar wind intervals exhibit the $k^{-5/3}$ and $k^{-2.8}$ spectral slopes at inertial and ion-kinetic scales, but the fast solar wind typically had more energy in the inertial range, offsetting the spectra from one another \citep{Alexandrova:2009}. Other fast-slow solar wind interval pairs taken on either side of high speed streams and over a wider range of plasma conditions showed the same spectral slopes behavior but without the offset \citep{Bruno:2017}. This universality of spectral slopes can persist across the interplanetary shock into the CME sheath region \citep{Bourouaine:2012, Good:2020}, and can be seen in the WIND data of Figures~\ref{fig:psd} and \ref{fig:psd-xyz}. As the sheath evolves from L1 to Earth, the ion-kinetic range spectral slope steepens so that the turbulence departs from this universal energy transfer rate, but the total energy in the inertial range does not increase except between the spectral breakpoints at L1 and Earth. This increase in slope is associated with an increase in intermittency due to a decrease in the probability of medium-sized fluctuations, particularly from inertial to the ion-kinetic scales (Fig.~\ref{fig:intermittency-xyz}). It seems as though the turbulence favors large amplitude fluctuations in the sense that the energy in the medium size fluctuations is more readily dissipated while the energy in large-amplitude fluctuations is not. Since the magnetic compressibility is $< 2$, at first glance these fluctuations are Alfvenic and not current sheets. The medium size fluctuations disappear, making the magnetic field more smooth and increasing the correlation length. The loss of medium-sized structures that occurs between L1 and Earth is associated with a departure from the universal background cascade of the solar wind and an increased energy cascade in the CME sheath.

The radial evolution of the spectral slopes and spectral breakpoints has also been studied in the solar wind \citep{Bourouaine:2012, Lotz:2023}. The spacecraft-frame spectral break frequency observed by Parker Solar Probe falls as $r^{-1.1}$ from $r=0.1-0.65\,\mathrm{R_{E}}$ \citep{Lotz:2023} but the doppler-shifted frequency, as observed by Helios 2, is roughly constant as a function of $r$ \citep{Bourouaine:2012}. The spectral index observed by PSP is relatively constant at -1.6 in the inertial range, while the index in the ion kinetic range becomes shallower, decreasing toward -2.8 near $1\,\mathrm{AU}$ \citep{Lotz:2023}. In addition, negative solar wind energy cascade rates at MHD scales have been reported at approximately $1\,\mathrm{AU}$ and at larger heliocentric distances \citep{Smith:2009, Hadid:2017, Stawarz:2010, Romanelli:2024}. Our observations of a CME sheath propagating from L1 to Earth show a constant inertial range spectral index of -1.6, an ion kinetic range spectral index that steepens from -2.8 to -3.6, and a breakpoint that shifts to smaller scales, with energy appearing to flow toward larger scales to compensate. Given the Alfv\'enic nature of the turbulence and the loss of intermediate amplitude fluctuations at ion scales, future work could check to see if wave damping is increasing the correlation of plasma and magnetic field fluctuations (cross-helicity); this could explain the steepened spectral slope and flow of magnetic energy to larger scales \citep{Smith:2009, Bourouaine:2012}.

Interplanetary shock and their accompanying sheaths are one of the most, if not the most, geoeffective events \citep{Lugaz:2016}, inducing intense geomagnetic storms. Detecting them and predicting their arrival times is an important focus of Space Weather forecasting \citep{Vorotnikov:2008} and highlights the importance of data from upstream monitors at the Earth-sun Lagrange 1 (L1) point. Unfortunately, single spacecraft data from L1, including the popular OMNI dataset, are not always representative of the conditions at Earth \citep{Nykyri:2019, Borovsky:2018, Burkholder:2020}. This is often attributed to the size of a flux tube or spatial gradients in the solar wind \citep{Borovsky:2018, Burkholder:2020} rather than temporal evolution of solar wind structures. Geoeffective properties of shock and sheath are related to the creation, compression, or pile-up of southward-directed magnetic field \citep{Lugaz:2016}. The erosion of magnetic flux draped around the magnetic ejecta can reduce geoeffectiveness. An open question is whether or not increased energy cascade or the loss of intermediate-sized fluctuations has an effect on geoeffectiveness.

The sheath region contains several structures that are highly geoeffective. Compression of pre-existing southward magnetic field ($B_{z} < 0$) behind the shock is visible in the first half of the sheath (Fig.\ref{fig:overview}). This is immediately followed by a potentially planar structure, granted with $B_{z} > 0$. Finally, there is evidence of field-line draping in front of the magnetic cloud, with $B_{z}$ rotating from positive to negative, a slight deflection of the bulk velocity, and an increased temperature. These large-scale structures are present at both L1 and Earth. The Pearson correlation coefficient between Wind and MMS of the time-shifted magnetic field components is $\rho > 0.93$, indicating that for this event the conditions observed by the solar wind monitor are representative of what was eventually observed by MMS despite the large dawn-dusk separation. However, at MMS, the turbulent cascade rate at ion kinetic scales increased (Fig.~\ref{fig:psd-xyz}) and the probability of intermediate-sized fluctuations decreased (Fig.~\ref{fig:intermittency-xyz}), making the sheath field smoother (Fig.~\ref{fig:corr-xyz}). Future modeling studies will be needed to determine if this has any significant impact on geoeffectiveness. 

\section{Conclusions}
We study the evolution of turbulence within the sheath region between an interplanetary shock and the associated magnetic cloud from L1 to Earth. While the MHD-like spectral slope in the inertial scale does not change, the spectral breakpoint shifts toward smaller spatial scales, closer to the ion inertial length, and the sub-ion-scale spectral slope steepens from $k^{-2.8}$ to $k^{-3.6}$. The shift in breakpoints corresponds to an increase in energy at the associated spatial scales and may indicate that reverse cascade occurs. The steepened spectral slope is confirmed by multi-spacecraft estimates of the equivalent spectrum. The spectral breakpoints and slopes present in the PSD of $B_{x}$ and $B_{z}$ evolve in a similar manner to those of the omni-directional PSD, but those of $B_{y}$ do not change from L1 to Earth, indicating that the turbulence evolves anisotropically.

These findings are supported by our analysis of intermittency and correlation length. $B_{x}$ and $B_{z}$ become more intermittent from L1 to Earth and from inertial- to ion-scales. At Earth, the fluctuations become more intermittent below ion-scales. Increased intermittency, however, is not due to an increase in the number of large-amplitude small-scale fluctuations. Instead, it is due to a reduction of intermediate ($\Delta B_{i}^{l} < 4$) fluctuations. Intermittency of $B_{y}$ stays roughly constant from L1 to Earth and does not exhibit an increase in intermittency below ion-scales, consistent with the PSD of $B_{y}$ not evolving from L1 to Earth. The turbulence observed at both L1 and Earth is highly Alfv\`enic in nature. The auto-correlation of $B_{y}$ follows closely what one would expect from the frozen-in flow conditions, and the associated correlation length is much shorter than that of $B_{x}$ and $B_{z}$. For $B_{x}$ and $B_{z}$, $\lambda_{C}$ increases from L1 to Earth, indicating that adjacent parcels of magnetic field become more similar. This reinforces the result that turbulence is anisotropic, energy cascade increases, and intermediate fluctuations decrease, indicating that CME sheath turbulence can evolve noticeably from L1 to Earth.

Future studies could compare the upstream solar wind, post-shock, pre-ejecta regions, and identify planar structures to better understand the sheath evolution \citep[e.g.,][]{Kilpua:2020}.
Another potential study could characterize the turbulent fluctuations \citep[e.g.,][]{ChenCHK:2017} to determine a source for and scale size at which the ion kinetic scale turbulence deviates from the universal background, and the probability of intermediate amplitude fluctuations decreases. 
Finally, one could analyze the Alfvénic nature of the fluctuations by combining plasma moments with the magnetic field data \citep[e.g.,][]{Stawarz:2010, Romanelli:2024} to investigate the transfer of magnetic energy to larger scales between L1 and Earth,.

\begin{acknowledgments}
This work is supported by NASA's MMS mission under contract NNG04EB99C. M.R.A. was supported by NASA H‐GIO Grant 80NSSC22K0187. 
N.R. is supported through a cooperative agreement with Center for Research and Exploration in Space Sciences \& Technology II (CRESST II) between NASA Goddard Space Flight Center and University of Maryland College Park  under award number 80GSFC24M0006. N.L. acknowledges support from NASA LWS grant 80NSSC24K1245. MMS data \citep{Burch:2024} is publicly available at the MMS Science Data Center 
(\hyperlink{https://lasp.colorado.edu/mms/sdc/public/}{https://lasp.colorado.edu/mms/sdc/public/});
it was accessed and analyzed using the PyMMS Python package \citep{Argall:2020a}. Wind data is publicly available through the Coordinated Data Analysis Web
(CDAWeb; \hyperlink{https://cdaweb.gsfc.nasa.gov/istp_public/}{https://cdaweb.gsfc.nasa.gov/istp\_public/})
and was accessed using Python via NASA's Coordinated Data Analysis System Web Service Client Library (CDASWS, \hyperlink{https://pypi.org/project/cdasws/}{https://pypi.org/project/cdasws/}).

\end{acknowledgments}

\bibliography{references}{}
\bibliographystyle{aasjournal}



\end{document}